\documentclass[twocolumn,floatfix]{revtex4-2}
\usepackage{amsmath}
\usepackage{graphicx}
\usepackage{epsfig}
\usepackage{color}
\usepackage{soul}
\usepackage{longtable}
\usepackage{xcolor}
\usepackage{cancel}
\usepackage{txfonts}
\usepackage{cancel}
\usepackage[normalem]{ulem}
\usepackage[utf8]{inputenc}
\usepackage{amssymb}
\usepackage{hyphenat}
\usepackage{hyperref}
\hypersetup{colorlinks=true, citecolor=blue, urlcolor=blue, linkcolor=blue}

\begin{document}

\title{Self Tuned Criticality: Controlling a neuron near its bifurcation point via temporal correlations}

\author{Juliane T. Moraes$^1$}
\email{juliane.moraes@ufv.br}
\author{Eyisto J. Aguilar Trejo$^{2,3}$}
\author{Sabrina Camargo$^{2,3}$}
\email{scamargo@unsam.edu.ar}
\author{Silvio C. Ferreira$^{1,4}$}
\author{Dante R. Chialvo$^{2,3,5}$}

\affiliation{
$^1$ Departamento de F\'isica, Universidade Federal de Viçosa, Minas Gerais, Brazil\\
$^2$ Center for Complex Systems and Brain Sciences
(CEMSC3),Instituto de Ciencias F\'isicas (ICIFI), Universidad Nacional de San Mart\'{\i}n, Campus Miguelete, San Mart\'{\i}n,
Buenos Aires, Argentina.\\
$^3$ Consejo Nacional de Investigaciones Cient\'{\i}fcas y Tecnol\'ogicas (CONICET), Buenos Aires, Argentina.\\
$^4$
National Institute of Science and Technology for Complex Systems, 22290-180, Rio de Janeiro, Brazil.\\
$^5$
Mark Kac Center for Complex Systems Research and Institute for Theoretical Physics, Jagiellonian University, Kraków, Poland.}
 \date{\today}

\begin{abstract}
Previous work showed that the collective activity of large neuronal networks can be tamed to remain near its critical point by a feedback control that maximizes the temporal correlations of the mean-field fluctuations.
Since such correlations behave similarly near instabilities across nonlinear dynamical systems, it is expected that the principle should control also low dimensional dynamical systems exhibiting continuous or discontinuous bifurcations from fixed points to limit cycles. Here we present numerical evidence that the dynamics of a single neuron can be controlled in the vicinity of its bifurcation point. The approach is tested in two models: a 2D generic excitable map and the paradigmatic FitzHugh-Nagumo neuron model. The results show that in both cases, the system can be self-tuned to its bifurcation point by modifying the control parameter according to the first coefficient of the autocorrelation function. 

\end{abstract}

\maketitle

\emph{Introduction:} Experimental evidence \cite{beggs2003,haimovici2013}, as well as theoretical developments \cite{chialvo2010,mora2011,eguiluz2005}, suggest that the highly variable spontaneous activity of neuronal networks corresponds to critical phenomena. Current efforts are dedicated to better understanding the behavior of neurons and networks near that critical region. To track this transition, the divergence of the fluctuations (e.g., susceptibility) of the order parameter often is used \cite{marro2005}. Alternatively, in Ref. \cite{chialvo2020} the autocorrelation coefficient of the order parameter, which peaks at the critical point \cite{hohenberg1977} more smoothly than the susceptibility was used to control the system near the transition. In this work, we extend this control strategy to a much simpler scenario, namely a generic excitable dynamic modeled by low-dimensional systems.
    
The two models selected in the present work belong to a large variety of attempts made to describe the activity of single neurons \cite{izhikevich2007}. From intricate and realistic ionic models \cite{hodgkin1952} to simpler ones, all aim to reproduce the physiological states of neurons in terms of their electrical properties to understand and predict their behavior in response to perturbations. Here we limit ourselves to exploring two models: the first one is a two-dimensional difference equation (so-called Chialvo map), proposed for simulating generic neural dynamics \cite{chialvo1995}. The second one is a differential equation, the FitzHugh-Nagumo system (FHN) \cite{fitzhugh1955,fitzhugh1961,nagumo1962}. The behavior in parameter space of both models is very well known; they exhibit bifurcations from fixed points to limit cycles and to more complex behavior, such as chaotic bursting, as their parameters are varied. Both have an activation (potential-like) variable and a recovery-like variable. 

Previous work has shown the possibility of shifting the system to the critical point employing a feedback mechanism \cite{chialvo2020}. This method uses the first coefficient of the standard temporal autocorrelation function to modify the control parameter, driving and maintaining the system near the critical point. Low-dimensional models mentioned above  are known to exhibit a bifurcation rather than a phase transition. In this case, a vanishing small noise and a constant additive term over the activation variable allow to estimate the typical slowing down of the dynamic and thus allowing the control of the system in the vicinity of its bifurcation point. 
    
The paper is organized as follows: in the next section \ref{sec:chialvomap} we demonstrate the control mechanism in the Chialvo neuron map under two of its regimes: the excitable and the chaotic dynamics. Next,  the same procedure demonstrates the control of the excitable regime in the FitzHugh-Nagumo system. We close with a brief comment on the relevance of our results. \\

\begin{figure*}[ht] 
	\centering
	\includegraphics[width=.9\textwidth]{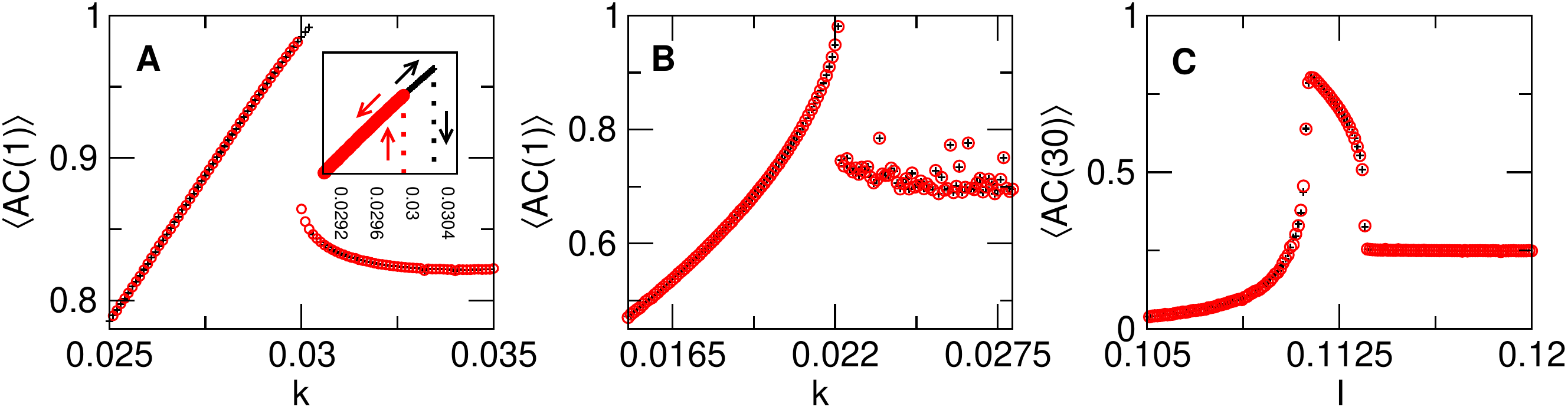}
	\caption{Average autocorrelation coefficient as a function of the control parameter for the Chialvo's neuron map in the excitable (panel A) and chaotic (panel B) regimes, for one hundred samples. The inset in panel A depicts the typical hysteresis for the excitable regime of this model.  Panel C shows the same quantities for the FHN model. In all the panels, the evolution of $AC$ is shown for increasing (black crosses) and decreasing (red circles) control parameter values.}  
	\label{fig:equilibrium}	
\end{figure*}

\emph{The Chialvo map:} \label{sec:chialvomap} In physics, the use of maps (i.e., difference equations), instead of differential equations, was championed by Kaneko decades ago \cite{kanekobook1,kaneko2} when he argued that a great variety of complex spatio-temporal phenomenology was not dependent on the precise and detailed account of the local dynamics but rather on the interaction of a large number of nonlinear degrees of freedoms. The approach, known as coupled-map lattices, was used since then to study a variety of complex phenomena \cite{kaneko3,kanekobook2}. It was natural to attempt the modeling of neuronal activity with maps, because the approach may provide a simple analytical framework to study its bifurcations and chaotic orbits, exact solutions, etc., and at the same time considerably speed up the computations of large neuronal networks. One of the first models, introduced by Chialvo \cite{chialvo1995} two decades ago, is a two-dimensional map capturing the most generic excitable dynamic including relevant physiological properties e.g., the behavior of the resting and the membrane action potential, its refractory period, and the periodic oscillations under constant bias input that may bifurcate to chaotic solutions (see reviews of similar approaches in Refs. \cite{girardi2013,ibarz,Courbage,Dmitrichev}). The model is a two-dimensional difference equation 
\begin{equation}
\begin{array}{rl}
x_{n+1} =& f(x_n,y_n) = x_n^2 \exp{(y_n-x_n)}+k\\[0.5pt]
y_{n+1} =& g(x_n,y_n) = ay_n-bx_n+c    
\end{array}
\label{eq:chialvo_map}
\end{equation}
in which $n$ is an iteration step, $x$ represents the activation variable, $y$ acts as a recovery-like variable, $a$ is the time constant of the recovery ($a<1$) variable, $b$ is the activation dependence of the recovery process ($b<1$), and $c$ is a constant which sets the fixed point of the recovery variable. The parameter $k$ can represent a constant bias or a time-dependent additive perturbation and will act as the control parameter.  
A compact view of the asymptotic behavior of any initial condition can be represented by the vector field in the phase plane of the $x$ and $y$ nullclines of the map, which are 
\begin{equation}
\begin{array}{rl}
x_f =& x_f^2 \exp{(y_f-x_f)}+k\\[0.8pt]
y_f =& \dfrac{c-bx_f}{(1-a)}    
\end{array}
\label{eq:fixed_points}
\end{equation}
 \begin{figure}[h!] 
	\centering
	  \includegraphics[width=.65\columnwidth]{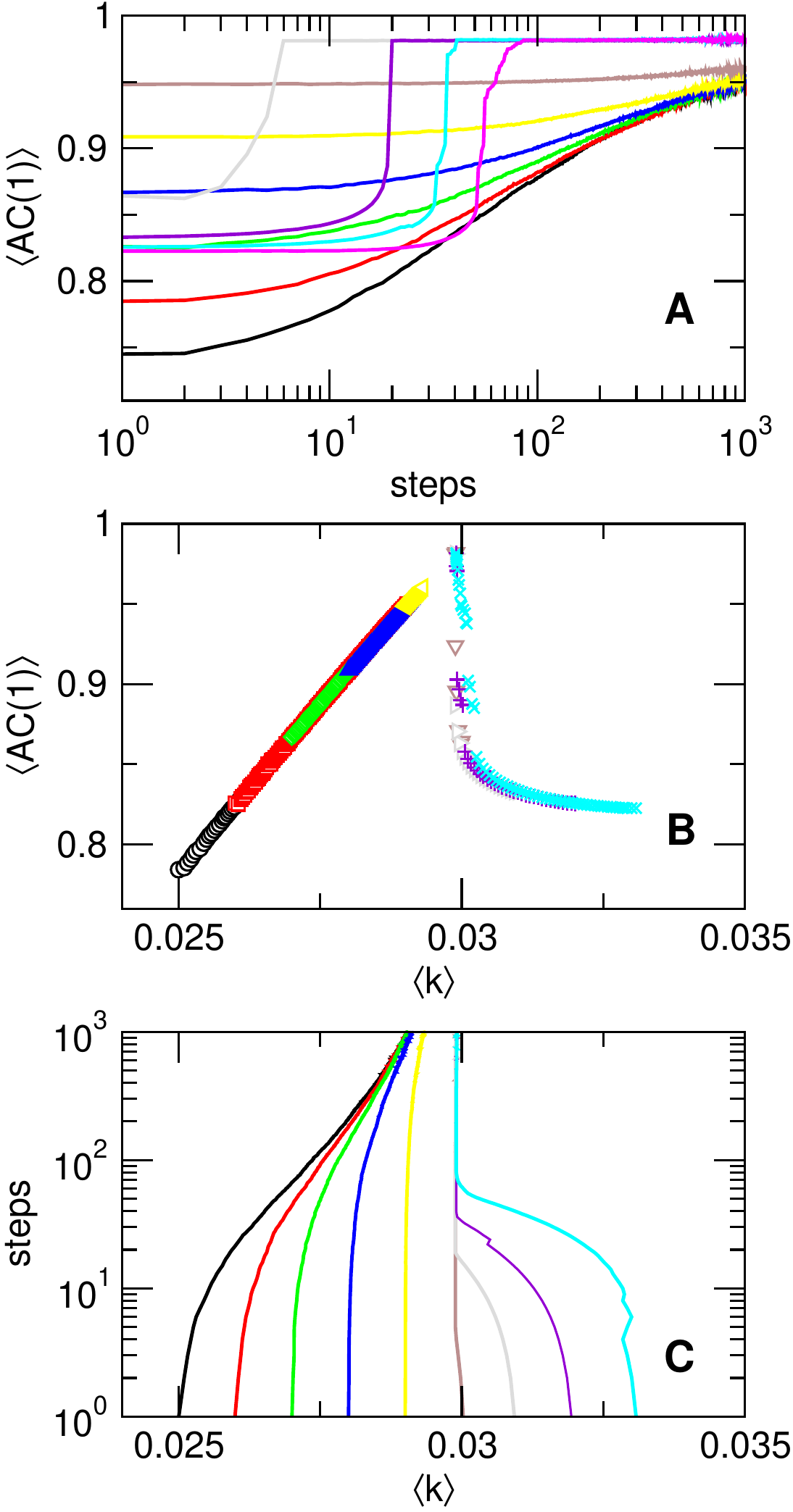}
  	\caption{Adaptive control of the Chialvo's neuron` map in the excitable regime. All of the results were averaged over one hundred samples. Data correspond to numerical solutions of the model starting from ten different initial conditions of the parameter $k$ from $0.024$ (black) to $0.033$ (magenta). Different colors denote the evolution of the variables toward the bifurcation point for each initial condition. Panels A and C show the evolution of $AC(1)$ and control parameter $k$ respectively. Panel B shows $AC(1)$ as a function of the control parameter $k$. Parameters $(a,b,c) = (0.89, 0.60, 0.28)$.} 
	\label{fig:map_excitable}	
\end{figure}
The control strategy discussed here relies on the behavior of the system's fluctuations near instabilities. Near such points, the amplitude of the fluctuations typically increases, and its temporal correlations slow down\cite{meisel}.  Because we consider here a deterministic single neuron model by itself devoid of any fluctuation, it becomes necessary to add a vanishing small Gaussian noise term to the activation variable. When the system is far from the bifurcation point, these perturbations will decay fast, resulting in a relatively low autocorrelation value. Conversely, as the system moves to the unstable regime, there is the typical critical slow-down of the perturbations and the auto-correlation grows. An estimator of this behavior is the autocorrelation function computed from a short time series of a system variable of interest. Here we used the value of the autocorrelation function at a sufficiently short lag, usually termed AC(1).

The data in Figure \ref{fig:equilibrium} shows the variation of  AC(1) as a function of the bias for Chialvo's map and FHN model. Notice that in the three examples, the value of the autocorrelation decays as the control parameter moves away from the critical point, 
a feature that is expected from the theory (see, for instance a recent review \cite{grigera}, as well as  some relevant examples in Refs.\cite{chialvo2020, tricritical, box-scaling, sanchez}). 
%
\begin{figure*}[ht!]
    \centering
    \includegraphics[width=.5\textwidth]{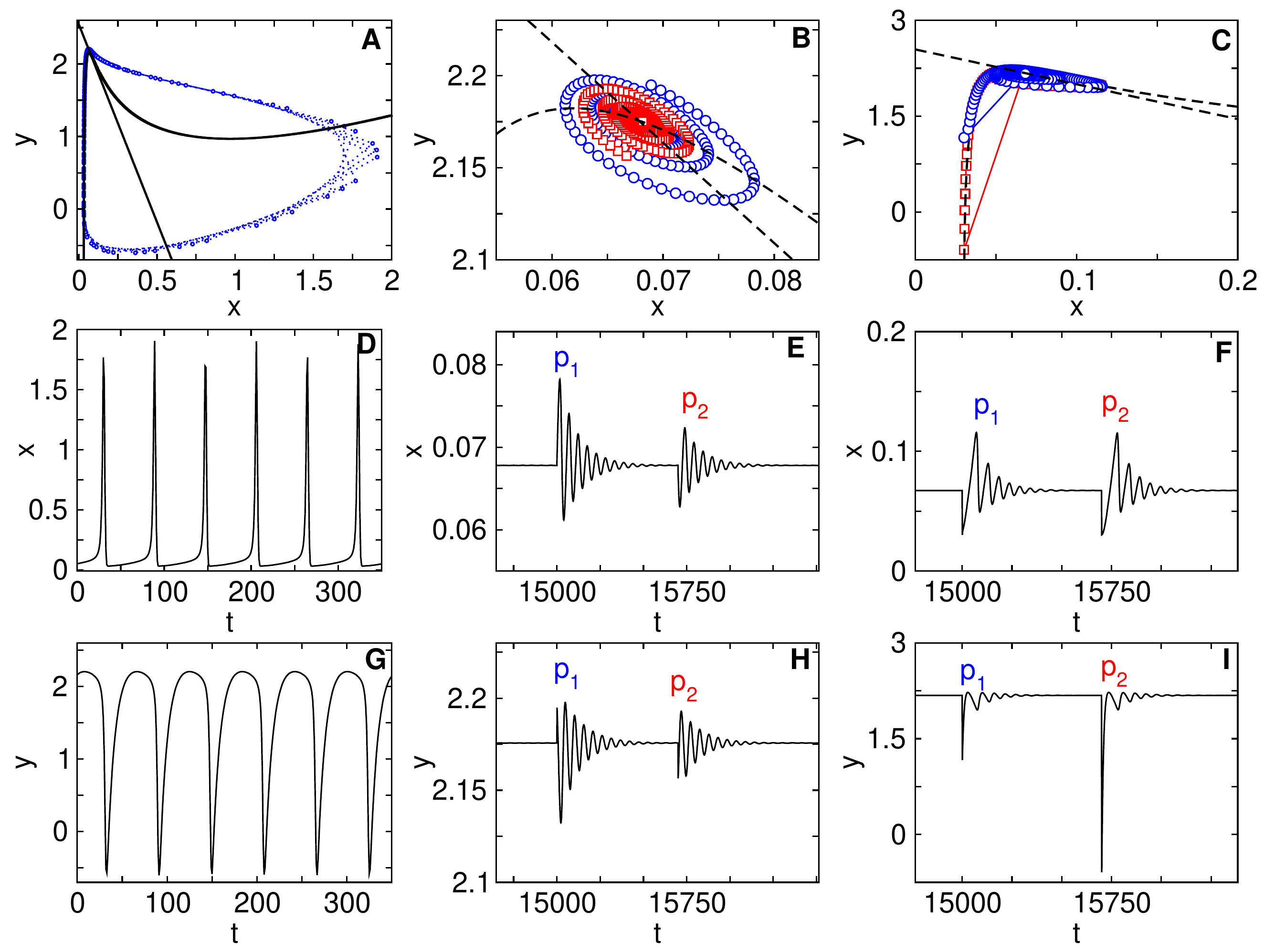}
    \caption{Adaptive control of the Chialvo neuron map in the excitable regime. Panel A shows the evolution of the two variables (blue dotted lines) and the nullclines of Eq.~(\ref{eq:fixed_points}) (black solid lines) without control (fixed $k=0.031$)  corresponding to the data in panels D and G. Panel B and C show the evolution and the nullclines after the control is switch on and after two perturbations (panels E and H,  and panels F and I respectively).  }
    \label{fig:map_excitable_traces}
\end{figure*}
Note the hysteresis (inset in panel A) already described \cite{chialvo1995} for the excitable regime of the Chialvo map. For the three cases to be described  the behavior of AC(1) as a function of the control parameter is similar, with some peculiarities to be discussed later.
According to the previous work of Ref.  \cite{chialvo2020}, control can be implemented via the parameter $k$ in Eq.~(\ref{eq:chialvo_map}), which could correspond to the neuron’s threshold  or the number of postsynaptic receptors, controlling the current it receives from other neurons. The procedure starts by selecting an initial random value of $k$ and simulating the dynamics of Eq.~(\ref{eq:chialvo_map}) for a large number of iterations. After that, we estimate the average value of AC(1) of the activation variable $x_n$ during these iterations. We then compare if the AC(1) value at the control step $i$ with the one computed at the previous step $i-1$ (see Eq.~(\ref{eq:control})), which determines the sign of $d_i$.  Finally, the gradient to the maximum (ie., AC=1) value $\delta_i$ is computed and used to adjust $k$ for the next control step. As the process is repeated, eventually, the gradient to AC=1  vanishes. 
\begin{equation}
    \begin{array}{rl}
         d_i = & d_{i-1} sign[AC_{(i)}-AC_{(i-1)}] \\
         \delta_i = & (1-AC_{(i)})^2 \\
         k_{i+1} = & k_i + \delta_i  \:  .   \: d_i  \:  .  \:  \kappa
    \end{array}
    \label{eq:control}
\end{equation}
More details about the control method in other models can be consulted in Ref.  \cite{chialvo2020}.
For the maps studied here, we use the value of the AC function at the shortest non-zero lag, usually termed AC(1). For the FHN model, we arbitrarily take the 30th lag given the fact that the FHN is a differential equation which offers a finer time resolution. The constant $\kappa$ is a small constant ($\kappa=0.0025$) that determines the time scale at which the control parameter is adjusted. Note that, as mentioned above, the autocorrelation measured represents the time scale of the model dynamic in response to a very small Gaussian perturbation (variance $\sigma \sim 10^{-6}$ for the map model and $\sigma \sim 10^{-5}$ for the FHN model).  \\
 \begin{figure}[hb!]
	\centering
	\includegraphics[width=.6\columnwidth]{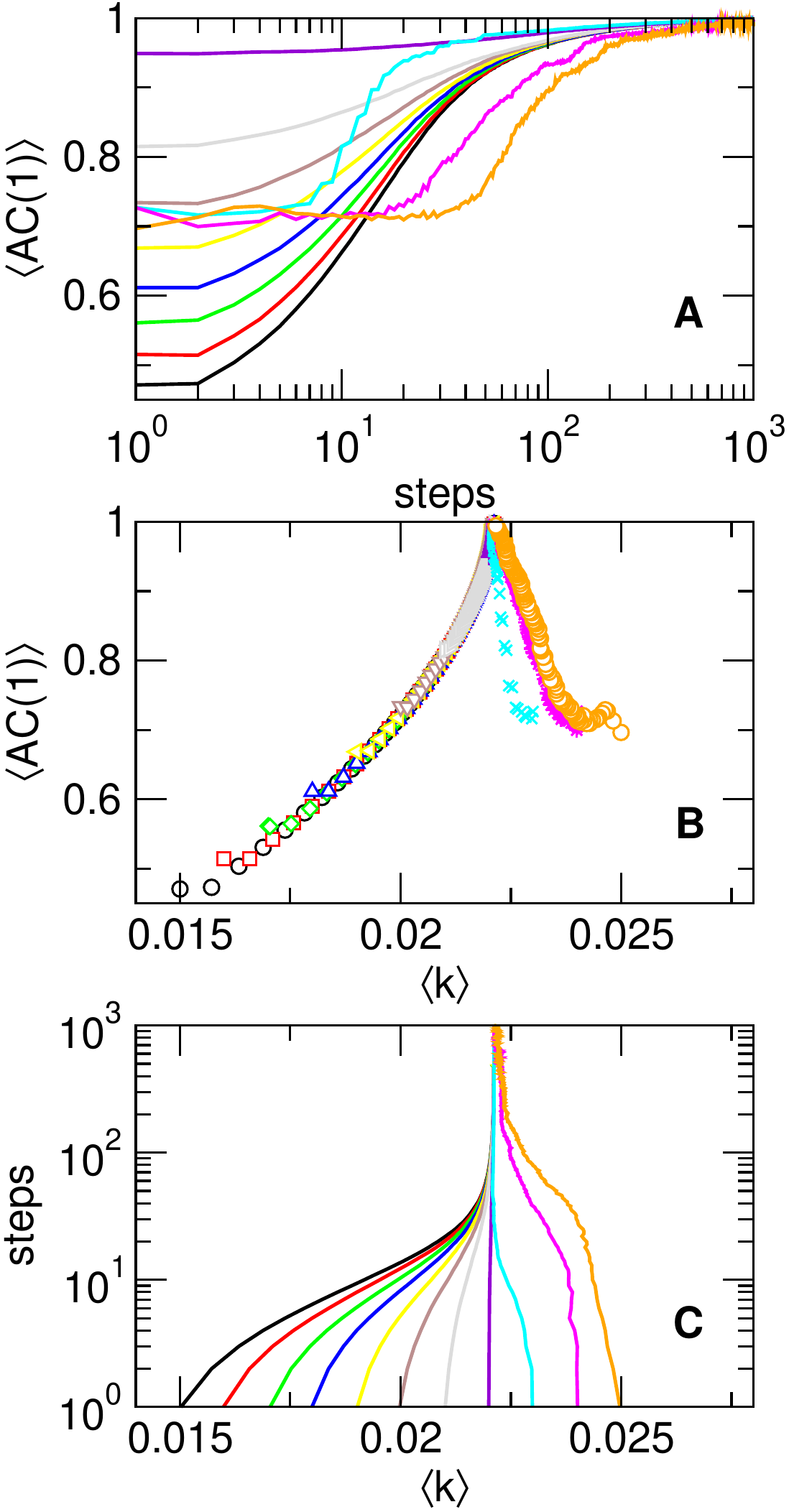}
	\caption{Adaptive control of the Chialvo's neuron map in the chaotic regime. Same format as in  Fig. \ref{fig:map_excitable}. Numerical solutions of the model starting from ten different initial conditions of the parameter $k$  from $0.015$ (black) to $0.025$ (orange), for 100 samples.}
	\label{fig:map_chaotic}	
\end{figure}

\begin{figure*}[ht!]
    \centering
\includegraphics[width=.5\textwidth]{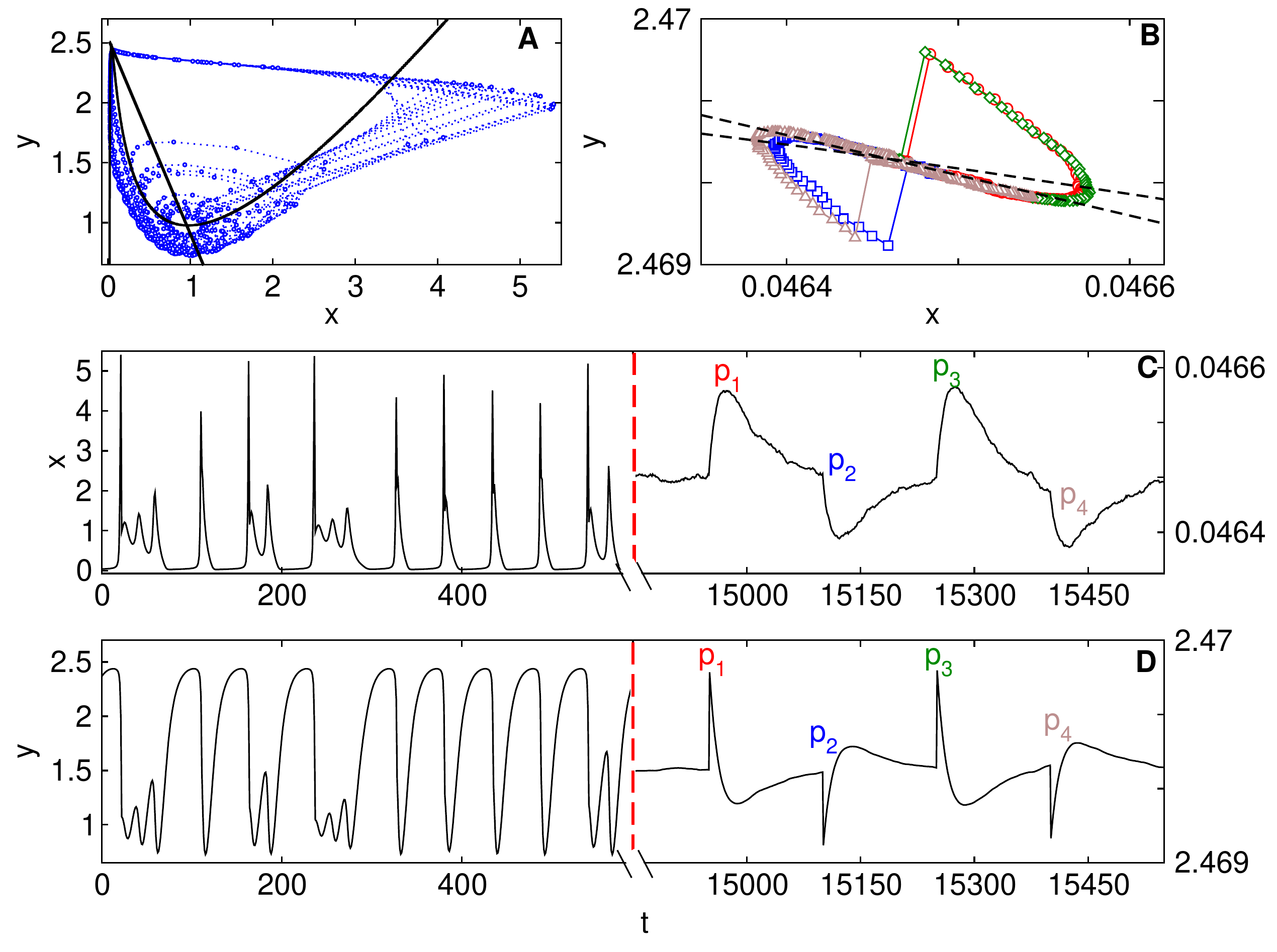}
    \caption{Example of the control applied to the variable $k$ in the Chialvo neuron map in the chaotic regime, with parameters $(a,b,c) = (0.89, 0.18, 0.28)$. In Panel A, the nullclines of Eq.~(\ref{eq:fixed_points}) are displayed by black lines while the blue dotted lines shows the trajectory for $k=0.025$. Trajectories in blue are the same displayed in the first half of panels C and D. Panel B shows the nullclines (dashed lines) and the symbols are the trajectory of the system after small perturbations, $p_1,p_2, p_3,$ and $p_4$ are applied to the system now under control, which can be seen in the second half of panels C and D. }
 \label{fig:map_chaotic_traces}
\end{figure*}
\emph{Chialvo map' excitable regime:} First, we explore the control of the map Eq.~(\ref{eq:chialvo_map}) for conditions in  which the system exhibits an excitable regime, corresponding to parameter values $(a,b,c) = (0.89, 0.60, 0.28)$. The control is performed  for $10^3$ adaptation steps, denoted with index $i$ in Eq.~(\ref{eq:control}) which in turn each adaptation lasts $4.10^4$ iterations of the map  Eq.~(\ref{eq:chialvo_map}) after discarding the transient ($3.10^4$ iterations). 

Figure \ref{fig:map_excitable} shows that for all cases the system converges to a region in which the control parameter remains fixed, and the autocorrelation achieves a maximum value. Notice, however, that instead of being a single asymptotic value for the control parameter there are two. This peculiarity is anticipated by the presence of hysteresis already mentioned in the description of the results in Figure \ref{fig:equilibrium}. For this model is known that for $k \approx 0.03$ there is bi-stability between a locally stable fixed point and periodic orbits  (see the example of Fig. 8 in Ref. \cite{chialvo1995}). The basin of attraction of the fixed point lies on a small region inside the orbit, while all initial conditions with values falling outside the orbit converge to the periodic solution. This implies, in the context of the control, that depending on the history (i.e., the initial conditions) the control will converge to one or the other asymptotic control points. This is evidenced by the results in Figure  \ref{fig:map_excitable_traces}. Each of the three columns presents the results obtained with the same simulations in three panels: the activation and recovery variable in the middle and bottom panels as a function of time. The same data is plotted in the phase space $x$ versus $y$, over-imposed to the nullclines of Eq. (\ref{eq:fixed_points}) in the top panels.  On the leftmost panels, the figure illustrates (for $k=0.031$) the non-controlled dynamics and the two other panels show the results under the control.  
For the middle panels, two small perturbations are applied and for the rightmost panels, two large perturbations are inserted. We observe that in both cases the system is led back to the controlled trajectories.

\emph{Chialvo map' chaotic regime:} It is known that for certain regions of parameter space  Eq.~(\ref{eq:chialvo_map}) exhibit chaotic dynamics. For instance, for parameters $(a,b,c,k) = (0.89, 0.18, 0.28,0.025)$ the system shows chaotic bursting, characterized by aperiodic trajectories  spiraling out of a repellor located in the proximity of the (x=y=1) region in phase space. An example is shown in Fig. \ref{fig:map_chaotic_traces}. \\

\emph{The FitzHugh-Nagumo neuron model:}
\label{sec:fhn}
Now we turn to replicate the above results in the paradigmatic FitzHugh-Nagumo model \cite{fitzhugh1955,fitzhugh1961,nagumo1962}. This formulation was inspired by the Van der Pol nonlinear neon-bulb relaxation oscillator \cite{vanderpol} model,  and also mimics the  celebrated Hodgkin-Huxley equations of the neuron membrane \cite{hodgkin1952}.  The equations are   
\begin{equation}
\begin{array}{c}
     v_{t+1} = v_{t} + dt[v_{t}(v_{t} - a)(1 - v_{t}) - w_{t} + I + \eta]/\epsilon\\[0.8pt]
    w_{t+1} = w_{t} + dt(v_{t} - dw_{t} - b)   
\label{eq:fhn}
\end{array}
\end{equation}
in which $v$ represents the membrane potential, or also called fast variable and $w$ is the slow recovery-like and $\eta$ is a very small perturbation. In the present simulations we use parameter $a = 0.5$, $b = 0.15$,  $d = 1$, $dt = 0.0015$, $\epsilon = 0.005$. The parameter $I$ acts as the control parameter.

\begin{figure} [ht!]
	\centering
	\includegraphics[width=.6\columnwidth]{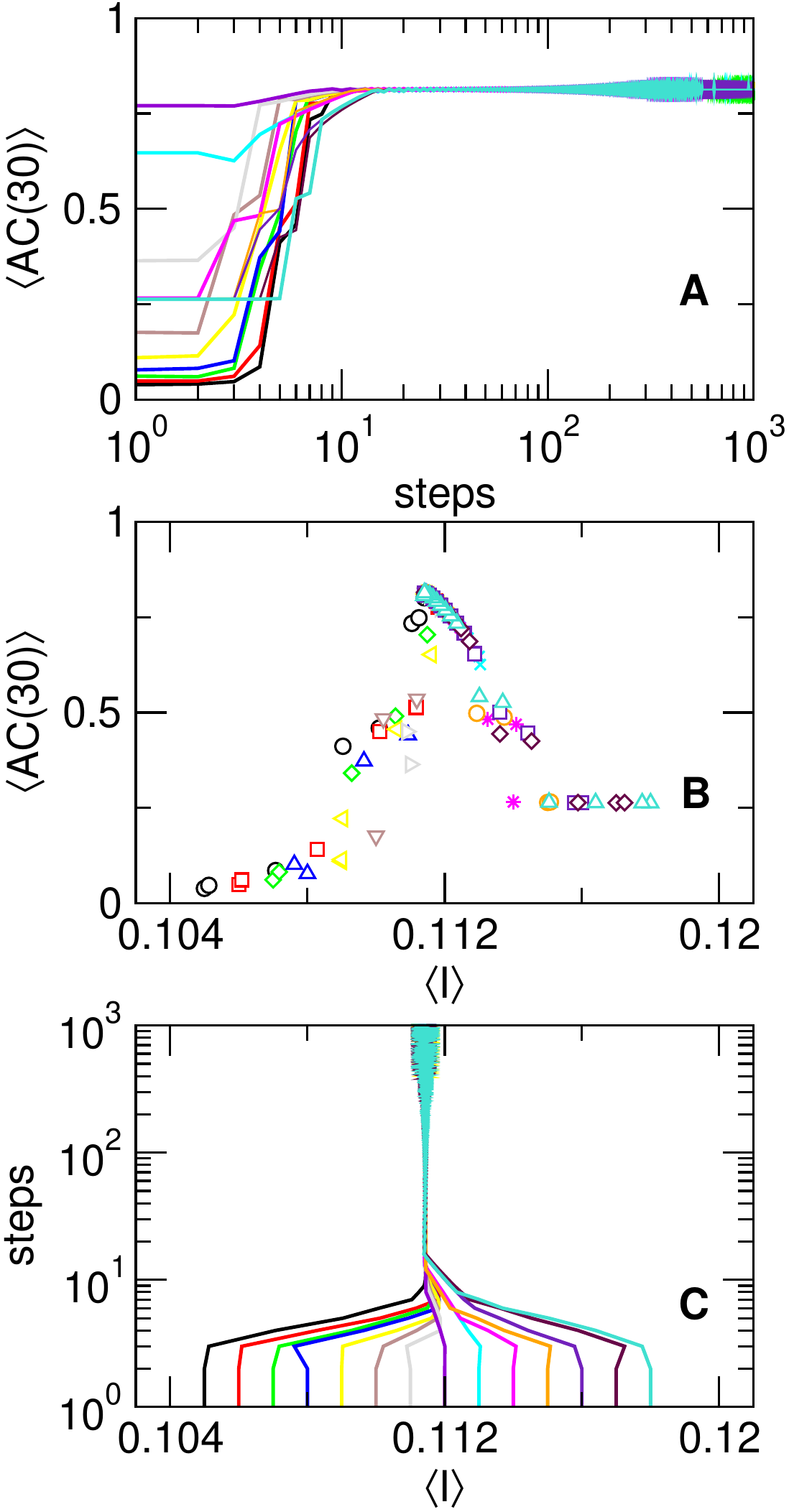}
	\caption{Adaptive control of the FitzHugh-Nagumo model. Numerical solutions of the model applying the control starting from fourteen different initial conditions of the control parameter $I$ (different colors), from $0.105$ (black) to $0.118$ (turquoise) for 100 samples.  } 
	\label{fig:FHN}	
\end{figure}
As shown in Figure \ref{fig:FHN}, the results for the control of the FitzHugh-Nagumo model are very similar to the Chialvo's map in the excitable regime. This is expected given the fact that for the parameter used here both models share the same bifurcation scenario; namely,  the FHN model undergoes a Hopf bifurcation from stable fixed point to a limit cycle oscillation as a function of a constant bias $I$.
The equations for the control are similar to Eq.~(\ref{eq:control}), which explicitly are 
\begin{equation}
    \begin{array}{rl}
         d_i = & d_{i-1} sign[AC(30)_{(i)}-AC(30)_{(i-1)}] \\
         \delta_i = & (1-AC(30)_{(i)})^2 \\
         I_{i+1} = & I_i + \delta_i  \:  .   \: d_i  \:  .  \:  \kappa
    \end{array}
    \label{eq:control_fhn}
\end{equation}
where $AC(30)$ is the autocorrelation coefficient (at lag=30) of the fluctuations of membrane potential $v$, using $\kappa=0.0025$.
In summary, the results show that the slowing down of the temporal correlations  generically present in a nonlinear system approaching an instability suffices to control and maintain it near such instability. Of course, this is not surprising, nonetheless, at first sight, it may appear paradoxical that the control is achieved at the point at which the system precisely is more  susceptible and consequently amplify  fluctuations the most. Therefore, our results may be potentially relevant to the work on homeostatic regulation of neuronal firing and plasticity\cite{turrigiano,hengen}. The present approach complements similar efforts directed to tune a system to its most sensitive working point, including early \cite{camalet,eguiluz}  and recent \cite{milewski} work in the cochlea. Other applications such as taming an ensemble of neurons monitored via optogenetics techniques \cite{Newman} in which it is possible to record and stimulate individual neurons. Meaning that this procedure can be a candidate to use the present control approach.

\emph{Acknowledgments:} Work conducted under the auspice of the Jagiellonian University-UNSAM Cooperation Agreement. Supported by the NIH BRAIN Initiative Grant No. 1U19NS107464-01, by CONICET and Escuela de Ciencia y Tecnolog\'ia, UNSAM, (Argentina) and by the Foundation for Polish Science (FNP) project TEAMNET ``Bio-inspired Artificial Neural Networks'' (POIR.04.04.00-00-14DE/18-00). The open-access publication of this article is supported in part by the program ``Excellence Initiative - Research University'' at the Jagiellonian University in Kraków (Poland), where part of this work was conducted.  J.T.M acknowledges the support and hospitality of the UNSAM where portion of this work was conducted as well the financial help of Coordenação de Aperfeiçoamento de Pessoal de N\'ivel Superior—CAPES - Finance Code 001, (Brazil).SCF thanks \textit{Conselho Nacional de Desenvolvimento Científico e Tecnológico} (CNPq)-Brazil (Grants no. 430768/2018-4 and 311183/2019-0) and \textit{Fundação de Amparo à Pesquisa do Estado de Minas Gerais} (FAPEMIG)-Brazil (Grant no. APQ-02393-18).

\end{document}